\documentclass[11pt,showpacs,preprintnumbers,superscriptaddress,amsmath,amssymb,nofootinbib]{revtex4}
\usepackage{graphicx}
\usepackage{dcolumn}
\usepackage{bm}
\usepackage{amssymb}
\usepackage{amsmath}
\usepackage{epsfig}    
\usepackage{color}
\usepackage{slashed}
\usepackage{hhline}
\usepackage{mathrsfs}

\def\be{\begin{equation}}
\def\ee{\end{equation}}
\def\Mat3#1#2#3#4#5#6#7#8#9{
  \left(
  \begin{array}{ccc}
    #1 & #2 & #3 \\
    #4 & #5 & #6 \\
    #7 & #8 & #9 \\
  \end{array}
  \right) }
\newcommand{\bea}{\begin{eqnarray}}
\newcommand{\eea}{\end{eqnarray}}
\newcommand{\nn}{\nonumber}

\numberwithin{equation}{section}

\usepackage{slashed,color,graphicx}
\usepackage{hyperref}
\hypersetup{
  colorlinks = true,
  linkcolor = blue,
  citecolor = magenta
}

\begin{document}

\title{Fermi-LAT GeV excess and muon $g-2$ in a modular $A_4$ symmetry }
\preprint{KIAS-P230xx, APCTP Pre2023-00x}
\author{Jongkuk Kim}
\email{jkkim@kias.re.kr}
\affiliation{School of Physics, KIAS, Seoul 02455, Korea}

\author{Hiroshi Okada}
\email{hiroshi.okada@apctp.org}
\affiliation{Asia Pacific Center for Theoretical Physics (APCTP) - Headquarters San 31, Hyoja-dong,
  Nam-gu, Pohang 790-784, Korea}
\affiliation{Department of Physics, Pohang University of Science and Technology, Pohang 37673, Republic of Korea}

\date{\today}

\begin{abstract}
The recent measurement of muon anomalous magnetic dipole moment (muon $g-2$) suggests that there might exist new physics  that dominantly interacts with muons.
The observed gamma-ray excess from Fermi-LAT indicates that dark matter annihilates into a specific charged fermions.  
We propose a successful model simultaneously to explain the Fermi-LAT GeV gamma-ray excess and sizable muon $g-2$ with a modular $A_4$ symmetry.
Due to nature of this symmetry, 
our DM only interacts with pairs of muon and 
we explain sizable muon $g-2$ without suffering from constraints of any lepton flavor violations.
We numerically show our allowed spaces on each measurements of Fermi-LAT, relic density of DM and muon $g-2$, randomly scanning our input parameters. 

\end{abstract}

\maketitle
\newpage

\section{Introduction}
Modular non-Abelian discrete flavor symmetries are widely applied to various flavor (new) physics such as quark and lepton Yukawa sectors to predict experimental values such as their masses, mixings, and phases as well as reproduce them, baryon asymmetry of Universe via leptogenesis, muon/electron anomalous magnetic dipole moment (muon/electron $g-2$), electron dipole moment (EDM), flavor changing processes like $\mu\to e\gamma$, $b\to s\ell\bar\ell$, and dark matter (DM). All of them (except quark and charged-lepton sectors) are expected to be existence of "beyond the standard model(BSM)".
Historically, this group was initially proposed by F. Feruglio in the paper~\cite{Feruglio:2017spp} and he investigated the lepton sector via a modular $A_4$ group. Subsequently, a large number of possibilities are arisen.   
For example, the following refs. are applied to lepton and quark sectors through the modular $A_4$ symmetry~\cite{Criado:2018thu, Kobayashi:2018scp, Okada:2018yrn, Okada:2019uoy, deAnda:2018ecu, Novichkov:2018yse, Nomura:2019yft, Okada:2019mjf, Ding:2019zxk, Nomura:2019lnr, Kobayashi:2019xvz, Zhang:2019ngf, Gui-JunDing:2019wap,
Kobayashi:2019gtp, Nomura:2019xsb, Wang:2019xbo, Okada:2020dmb, Okada:2020rjb, Behera:2020lpd, Nomura:2020opk, Nomura:2020cog, Asaka:2020tmo, Okada:2020ukr, Nagao:2020snm, Okada:2020brs, Kobayashi:2018vbk, Yao:2020qyy, Chen:2021zty, deMedeirosVarzielas:2021pug, Nomura:2021yjb, Ding:2021eva, Nagao:2021rio, Okada:2021aoi, Nomura:2021pld, Liu:2021gwa, Nomura:2022hxs, Otsuka:2022rak, Kang:2022psa, Ishiguro:2022pde,
Nomura:2022boj, Du:2022lij, Kikuchi:2022svo,
Gogoi:2022jwf, Abbas:2022slb, Petcov:2022fjf, Kikuchi:2023jap}.~\footnote{We restrict ourselves the refs. of $A_4$, even though various groups are adapted for flavor models.}
In addition to the lepton sector, leptogenesis is discussed in refs.~\cite{Kobayashi:2018wkl, Asaka:2019vev, Kashav:2021zir, Okada:2021qdf, Dasgupta:2021ggp, Behera:2020sfe, Ding:2022bzs}.
Semi-leptonic flavor changing neutral processes such as $b\to s\ell\bar\ell$, lepton flavor violations (LFVs), muon/electron $g-2$, and EDM are mainly discussed in refs.~\cite{Hutauruk:2020xtk, Kobayashi:2021pav, Kobayashi:2022jvy}.
Inflation model is also discussed in ref.~\cite{Gunji:2022xig}.
Interestingly, 
a successful DM candidate; i.e., stable DM, is realized by applying a modular symmetry.
Up to now, two methods are found.
 The first way is to apply a nonzero modular weight to DM that has a remnant symmetry such as $Z_2$ even after the modular symmetry breaking and the remnant one assures the stability of DM~\cite{Nomura:2019jxj}.
Another way is to make use of  
a recovering $Z_N$ symmetry that occurs at three fixed points of modulus $\tau$.
Here, we have $Z_2$ on $\tau=i$ and $Z_3$ on $\tau=\omega,\ i\infty$. Ref.~\cite{Kobayashi:2021ajl} realizes the stable DM on $\tau=i$ including the lepton sector where recovering symmetry is $Z_2$.~\footnote{The DM stability has been achieved by a way of a representation of traditional non-Abelian discrete flavor symmetries {\it e.g.}~\cite{Lavoura:2012cv, Hirsch:2010ru, Boucenna:2011tj}. Decaying DM with flavor specific case  has also been studied by these traditional symmetries~{\it e.g.}\cite{Kajiyama:2010sb, Haba:2010ag, Kajiyama:2013dba}.}

Recently, the Fermilab Muon $g-2$ Collaboration reported the anomalous muon magnetic moment \cite{Muong-2:2021ojo}.
The observation is consistent with the measurement by Brookhaven E821 Collaboration \cite{Muong-2:2006rrc}.
The $4.2\sigma$ deviation from the SM expectation may require new physics contributions to the muon $g-2$.
There have been comprehensive study on the context of the muon $g-2$ and other phenomena \cite{He:1990pn, Baek:2001kca, Chun:2019oix, Asai:2020qlp, Drees:2021rsg,Holst:2021lzm, Baek:2022ozm, Kim:2022xuo}.

From a decade ago, a gamma-ray excess at $E_\gamma\sim \mathcal{O}(1)$GeV coming from the Galactic center was found.
 In a lot of analyses from independent groups, it is claimed that the gamma-ray excess spectrum is in good agreement with the emission expected from DM annihilation into a specific charged particles.
The $b\bar{b}$ final state was understood by Higgs portal DM models and was studied \cite{Okada:2013bna,Ko:2014gha,Mondal:2014goi}. 
Also, the leptonic final state was understood by $U(1)_{L_\mu-L_\tau}$ model and investigated \cite{Park:2015gdo}.

In this paper, we explain the Fermi-LAT GeV excess that DM must annihilate into a pair of muons that produce gamma-ray to be observed.
The unique interaction of DM can be achieved by a modular $A_4$ symmetry, and
we apply the first way to stabilize our DM assigning nonzero modular weight.
As a bonus of our model, we also explain the sizable muon $g-2$ without worry about the LFVs such as $\mu\to e\gamma$ due to non-interactions except muon pair.

This paper is organized as follows.
In Sec.~\ref{model:setup}, we show our model setup starting from the relevant superpotential and soft breaking terms, and formulate the heavy charged lepton mass matrix.
In Sec.~\ref{DM:g2}, we firstly discuss our DM candidate on how the DM interact with the muon pair, and evaluate cross section requested by Fermi-LAT experiment as well as relic density of DM.
Then, we demonstrate the formula of muon $g-2$.
In Sec.~\ref{NumAnal}, we present our allowed region to  satisfy the Fermi-LAT measurement, relic density of DM, and muon $g-2$, randomly scanning our input parameters.  
In Sec.~\ref{cons}, we devote the conclusions and discussions. 

\begin{widetext}
  \begin{center} 
    \begin{table}
      \begin{footnotesize}
        \begin{tabular}{|c||c|c|c|c|c|c||c|c|c|c|c|c|}\hline\hline
          & \multicolumn{6}{c||}{SM leptonic superfields}& \multicolumn{5}{c|}{New superfields} \\\hline
            ~&~ $\widehat L_{e}$ ~&~ $\widehat L_{\mu}$ ~&~ $\widehat L_\tau$ ~&~ $\widehat {\overline e}$ ~&~ $\hat {\overline \mu}$ ~&~ $\widehat {\overline \tau}$ 
          ~&~ $\widehat L'$  ~&~ $\widehat {\overline {L'}}$ ~&~ $\widehat{E}$ ~&~ $\widehat{\overline E}$~ &~ $\widehat{\chi}$~ 
          \\\hline 
          $SU(2)_L$  & $\bm{2}$  & $\bm{2}$  & $\bm{2}$   & $\bm{1}$  & $\bm{1}$   & $\bm{1}$  & $\bm{2}$  & $\bm{2}$ & $\bm{1}$ & $\bm{1}$ & $\bm{1}$   \\\hline 
          $U(1)_Y$  & $-\frac{1}{2}$ & $-\frac12$ & $-\frac12$  & $1$ &  $1$  &  $1$  & $-\frac12$ & $\frac12$ & ${-1}$& ${1}$ & ${0}$    \\\hline
          $A_4$ & $1$  & $1$ & $1''$ & $1$  & $1''$   & $1'$ & $1$ & $1$ & $1'$ & $1''$  & $1$      \\\hline
          $-k$  & $0$  & $-2$ & $0$ & $0$ & $-2$ & $0$ & $-3$& $-1$ & $-3$& $-1$& $-3$     \\\hline
        \end{tabular}
        \caption{Field contents of the matter superfields
          and their charge assignments under $SU(2)_L\otimes U(1)_Y\otimes A_4$, where $-k$ is the number of modular weight.}
        \label{tab:1}
      \end{footnotesize}
    \end{table}
  \end{center}
\end{widetext}

 \section{Model setup} \label{model:setup}

 In this Section, we explain our
model construction, introducing the SM leptonic superfields and new ones and assigning the charges under the symmetries of  $SU(2)_L\otimes U(1)_Y\otimes A_4$, where $-k$ is the number of modular weight and "hat" over the fields represents superfields.
Here, we add one vector-like matter superfields
($\widehat E, \widehat{\overline E},\widehat L', \widehat{\overline {L'}}$) where we apply only fermionic parts of $\widehat E, \widehat{\overline E}, \widehat{L'}, \widehat{\overline{L'}}$ for our model.
 And $ E, \overline{E}$ are singly-charged fermions, while $L'\equiv[N',E']^T, \overline{L'}\equiv[\overline {N'},\overline{E'}]^T$ are isospin doublet ones.
 We use only a bosonic part of $\widehat\chi$ that is identified as a DM candidate, therefore $\chi$ is vanishing VEV.
$\chi$ is denoted by $\chi=(\chi_R + i\chi_I)/\sqrt2$ that is $A_4$ trivial singlet with $-3$ modular weight.
The DM $\chi$ also plays a role in inducing the muon $g-2$ together with $E$ and $L'$.
The superfield contents and their charged assignments are shown in Table~\ref{tab:1}.
%
Superfields $\widehat H_u$ and $\widehat H_d$ are introduced as MSSM.
We denote  their boson parts are written by $H_u=[h_u^+,(v_u+h_u+i z_u)/\sqrt2]^T$ and $H_d=[(v_d+h_d+i z_d)/\sqrt2,h_d^-]^T$ with totally neutral charges under $A_4$ and $(-k)$, where the structure is exactly the same as the minimum supersymmetric theory. The SM VEV is defined by $v_H\equiv \sqrt{v_u^2+v_d^2}\equiv 246$ GeV. 
Under these symmetries, the valid superpotential is found as follows:
\begin{align}
 \mathcal{W}_Y & = 
  y_e \widehat {\overline e} \widehat H_d\widehat L_e  +y_\mu \widehat {\overline \mu} \widehat H_d\widehat L_\mu  + y_\tau   \widehat{\overline\tau} \widehat H_d \widehat L_\tau
 + g_E \widehat {\overline {L'}} \widehat L_\mu \widehat\chi  + y_E \widehat{\overline\mu} \widehat E \widehat\chi
    \nn\\
  & h_E \widehat{\overline E}\widehat H_d\widehat L'+ M_E \widehat{\overline E} \widehat E   + M_{L'} \widehat {\overline {L'}} \widehat L' +\mu_H\widehat H_u\widehat H_d
  +\mu_\chi \widehat \chi\widehat \chi,
\label{yukawa}
\end{align}
where all couplings except $y_e,\ y_\tau,\ \mu_H$ implicitly includes modular Yukawa couplings that are determined only by modulus value. 
Charged-lepton sector does not mix each other, therefore the charged-lepton fields are mass eigenstates.
Due to even (odd) number of assignments for SM (new) fields in modular weight,
it is forbidden
for any mixing terms between SM and new fields such as $\widehat{\overline \ell} \widehat{H_d} \widehat{L'}$,
$\widehat{\overline{\ell}}\widehat{E}$, and $\widehat{\overline{L'}}\widehat{L_i}$ where $\ell=e,\mu,\tau$.
This is because any odd numbers of modular couplings are mathematically forbidden. However, we implicitly impose R-parity in order to prohibit superpotential such as
$\widehat{\overline e} \widehat{L_\mu}  \widehat{L_\mu}$,
$\widehat{L_e}  \widehat{H_u}$,
$\widehat{\overline e}\widehat{H_d}\widehat{H_d}$.
The bosonic mass of $\chi$ is obtained by the following soft SUSY breaking terms; 
\begin{align}
-{\cal L}_{\rm soft} \sim&  \mu_{B\chi}^2 \chi^2  + m^2_{\chi}|\chi|^2
+ {\rm h.c.}, \label{eq:pot}
\end{align}
where $m_\chi^2$ includes F-term.  Thus, mass of $\chi$ is given by
$m_{\chi_R}^2=  m_\chi^2 +  \mu_{B\chi}^2$
and $m_{\chi_I}^2= m_\chi^2 - \mu_{B\chi}^2$, and we select DM as $\chi\equiv \chi_I$, taking $0<\mu_{B\chi}^2$.

{\subsection{Heavy charged lepton mass matrix}
After the electroweak spontaneous symmetry breaking, a heavily charged lepton mass matrix is found {on the basis} of $[E,E']_L^T$ as 
\begin{align}
\mathcal{M}_E  = 
\begin{pmatrix}
M_E & m_E \\ 
0 & M_{L'} \\ 
\end{pmatrix},
\end{align}
where $m_E\equiv h_E v_d/\sqrt2$, and all the mass components are supposed to be real without loss of generality after rephasing of fields. 
Then, $\mathcal{M}_E$ is diagonalized by bi-unitary matrix as diag$[ m_{1}, m_{2}]=V^\dag_{E_R}  \mathcal{M}_E V_{E_L}$, where we define
\begin{align}
\begin{pmatrix}
E^\pm  \\ 
E'^\pm  \\ 
\end{pmatrix}
\equiv
\begin{pmatrix}
c_E & -s_E \\ 
s_E & c_E \\ 
\end{pmatrix}
\begin{pmatrix}
\psi_{E_1}^\pm \\ 
\psi_{E_2}^\pm  \\ 
\end{pmatrix}
,\label{massmat}
\end{align}
where $s_{E}(c_E)$ is short-hand notation for $\sin\theta_E(\cos\theta_E)$, and we abbreviate the electric charge hereafter.
Since these fermions have nonzero electric charges with specific decays, there exist lower bounds on masses.
If it decays into muon and missing, the bound is about 90 GeV~\cite{Workman:2022ynf}. 
In our case, $\psi_{E_1}^\pm$ decays (in limit of $s_E=0$) into muon and missing energy ($=$DM) via $y_E$ and $\psi_{E_2}^\pm$ decays (in limit of $s_E=0$) into muon, muon-neutrino, and the missing energy via kinetic term of $L'$ and $g_E$. We adopt the relaxing lower mass bound of 90 GeV in our numerical analysis. 
}

\section{Dark matter and muon anomalous magnetic dipole moment} \label{DM:g2}

\subsection{Dark Matter and Fermi-LAT GeV excess}
We consider a DM candidate $\chi_I$.
Then, the dominant contribution to the relic density of DM comes from the following relevant Lagrangian:
\begin{align}
-{\cal L} & = 
 \frac{i}{\sqrt2} g_E \bar {E'}  \ell_\mu \chi_I  + \frac{i}{\sqrt2} y_E  {\bar\mu}   E  \chi_I +{\rm c.c.}\nn\\
&= \frac{i}{\sqrt2} g_E (s_E \bar\psi_{E_1} +c_E \bar\psi_{E_2})  \ell_\mu \chi_I  + \frac{i}{\sqrt2} y_E  {\bar\mu}    (c_E \psi_{E_1} -s_E \psi_{E_2})  \chi_I +{\rm c.c.} ,
\label{yukawa}
\end{align}
where we suppose $g_E,\ y_E$ to be real without loss of generality.
%
%
Then, the {cross-section} is expanded by relative velocity $v_{\rm rel}$ that is given by $\sigma v_{\rm rel}\approx a_{\rm eff} + b_{\rm eff} v^2_{\rm rel}+{\cal O}(v^4_{\rm rel})$, where $a_{\rm eff}$ is s-wave and $b_{\rm eff}$ is p-wave.~\footnote{Even though we show this expansion up to p-wave, we have computed the relic density of DM up to d-wave that is proportional to $v^4_{\rm rel}$.} Each of waves is written by
\begin{align}
a_{\rm eff}&\approx \frac{(y_E g_E s_E c_E)^2}{2\pi}
\left[
\frac{m_1^2}{(m_\chi^2+m_1^2)^2} + \frac{m_2^2}{(m_\chi^2+m_2^2)^2}
\right], \\
b_{\rm eff}&\approx 
-\frac{(y_E g_E s_E c_E m_\chi)^2}{6\pi(m_\chi^2+m_1^2)^4(m_\chi^2+m_2^2)^4}
\left[
m_2^2(m_1^8 + 4 m_1^6 m_\chi^2+ m_\chi^8)(3 m_2^2+m_\chi^2)\right.\nn\\
&\left.\hspace{0.5cm}
+3 m_1^4(m_2^8+ 4m_2^6 m_\chi^2+12 m_2^4m_\chi^4+6m_2^2 m_\chi^6+m_\chi^8)\right.\nn\\
&\left.\hspace{0.5cm}
+ m_1^2 m_\chi^2(m_2^8+ 4m_2^6 m_\chi^2+18 m_2^4 m_\chi^4+8m_2^2 m_\chi^6+m_\chi^8)
\right],
\end{align}
where we define the DM mass to be $m_\chi$.
Since the cross section of Fermi-LAT $(\sigma v)_{\rm FL}$ is measured at the present Universe, $v_{\rm rel}$ is almost zero.
It leads us to be $a_{\rm eff}\approx(\sigma v)_{\rm FL}$. 
Then, one simply solves this equation in terms of one of our input parameters $s_E,\ y_E,\ g_E$.
Here, we solve it in term of $g_E$ and the result is straightforwardly given by
\begin{align}
g_E = \pm\sqrt{\frac{2(\sigma v)_{\rm FL}}
{(y_E  s_E c_E)^2
\left[ \frac{m_1^2}{(m_\chi^2+m_1^2)^2} + \frac{m_2^2}{(m_\chi^2+m_2^2)^2}\right] }}.
\label{eq:ge}
\end{align}
Thus, $g_E$ is not independent parameter any more hereafter, and we have to impose perturbative limit which we take $|g_E|\le1$ for more conservative manner.

To explain the gamma-ray GeV excess from galactic center, DM annihilation to the different SM particles is suggested. 
One of the solutions for the Fermi-LAT Gamma-ray excess is that the DM annihilation cross section to muons is $(3.9_{-0.6}^{+0.5})\times 10^{-26}$cm$^3$/sec with DM mass $58_{-9}^{+11}$ GeV at 1$\sigma$ interval \cite{DiMauro:2021qcf}.
The required DM annihilation cross section is very close to the canonical value for thermal freeze-out DM relic.

Relic density of DM is found in terms of expansion coefficients of  $v_{\rm rel}$ and 
approximately given by~\cite{Bertone:2004pz}
\begin{align}
\Omega_{\rm DM} h^2\approx \frac{1.07\times 10^9 x_F^2}{\sqrt{g_*}M_{\rm PL} (a_{\rm eff} x_F +3 b_{\rm eff} )},
\end{align}
where $M_{\rm PL}=1.22\times 10^{19}$ GeV, and $x_F\approx 24$.
In our numerical analysis, we apply the relaxed observed relic density $0.11\le \Omega_{\rm DM} h^2\le 0.13$ that is about 5$\sigma$ interval instead of exact value $\Omega_{\rm DM} h^2=0.1197\pm0.0022$ at 1$\sigma$ interval~\cite{Planck:2015fie}.

\subsection{Muon anomalous magnetic dipole moment: $\Delta a_\mu$} \label{lfv-lu}
A muon anomalous magnetic dipole moment ($\Delta a_\mu$ or muon $g-2$) has been {first} reported by Brookhaven National Laboratory (BNL) \cite{Muong-2:2006rrc}. They found that the muon $g-2$ data has a discrepancy at the 3.3$\sigma$ level from the SM prediction. Recent experimental result of muon $g-2$ suggests the following value at $4.2\sigma$~\cite{Muong-2:2021ojo}:
 \begin{align}
   \Delta a_\mu = a_\mu^{\rm{EXP}}-a_\mu^{\rm{SM}} = (25.1\pm 5.9)\times 10^{-10}.
   \label{eq:yeg2}
 \end{align}
%

To get sizable muon $g-2$ at one-loop level, we would need a chiral flip diagram and 
this contribution is obtained by the same term as Eq.(\ref{yukawa}).
Our muon $g-2$ at one-loop level is found as follows~\cite{Hutauruk:2020xtk}:
\begin{align}
&\Delta a_\mu
=
 \frac{m_\mu}{(4\pi)^2} y_E g_E s_E c_E 
 \left[F_I(m_\chi, m_{1}) - F_I(m_\chi,m_{2})\right],\\
& F_I(m_a,m_b)\simeq 
-\frac{m_b\left(3 m_a^4 - 4 m_a^2 m_b^2 + m_b^4 + 4 m_a^4 \ln\left[\frac{m_b}{m_a}\right]\right) }
{2(m_a^2-m_b^2)^3}.
\label{eq:damu1}
\end{align}

\section{Numerical analysis} \label{NumAnal}

In our numerical analysis, we randomly scan free parameters such that 
\begin{align}
\{y_E, s_E\} \in [0.01, 1], \quad \{m_1, m_2\} \in [90,1000]\ {\rm GeV}, \label{eq:num}
\end{align}
where the other parameter $g_E$ is determined by  Eq.~(\ref{eq:ge}).
Importing the measured observables $m_\chi,\ (\sigma v)_{\rm FL},\ \Omega_{\rm DM} h^2,\ \Delta a_\mu$, we plot two figures below where we take them up to 2$\sigma$ interval for $(\sigma v)_{\rm FL},\ \Delta a_\mu$, 1$\sigma$ interval for $m_\chi$,
and 5$\sigma$ for $\Omega_{\rm DM} h^2$ as discussed in the previous section.

\begin{figure}
\includegraphics[width=0.45\linewidth]{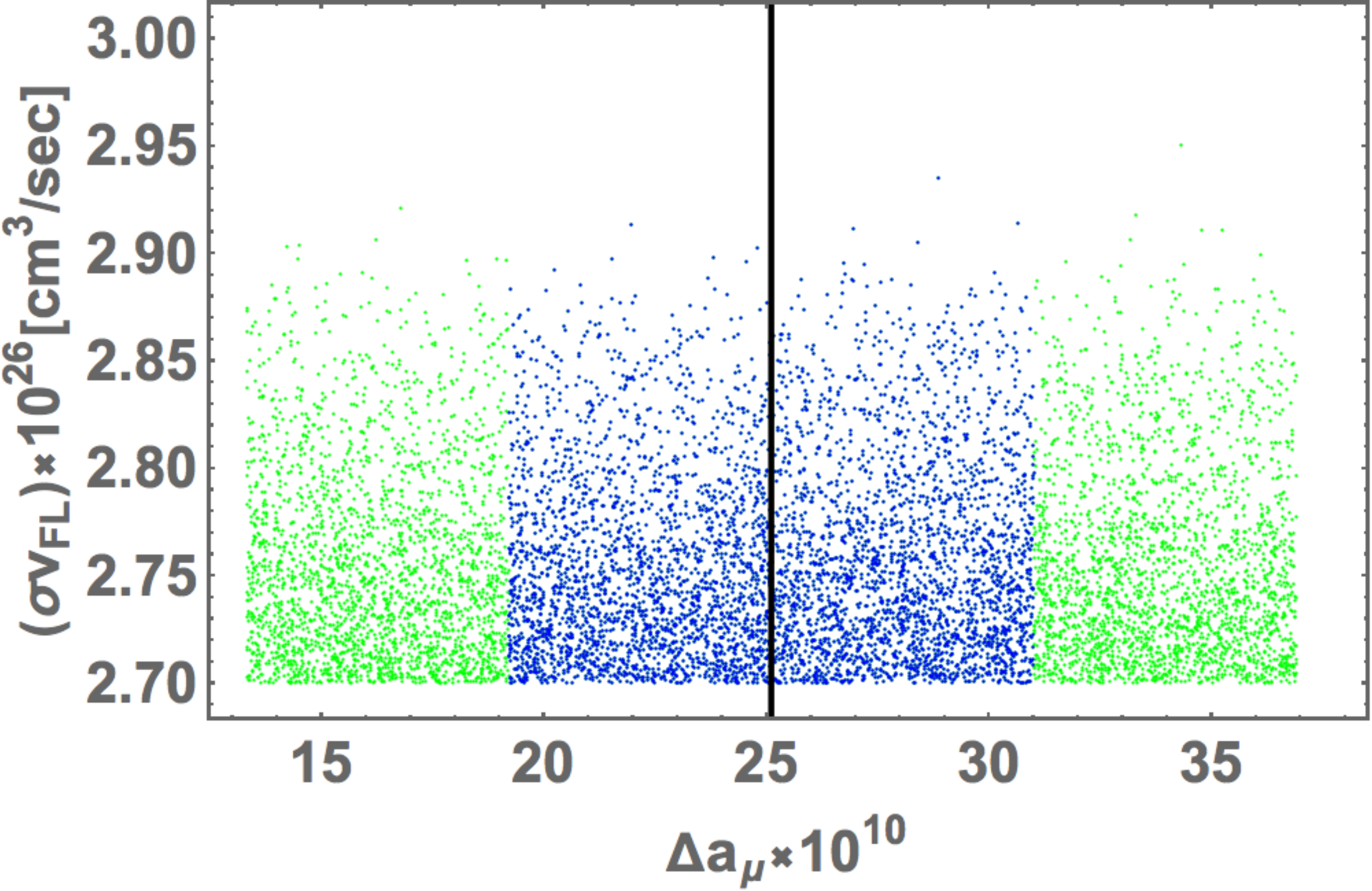}
\includegraphics[width=0.45\linewidth]{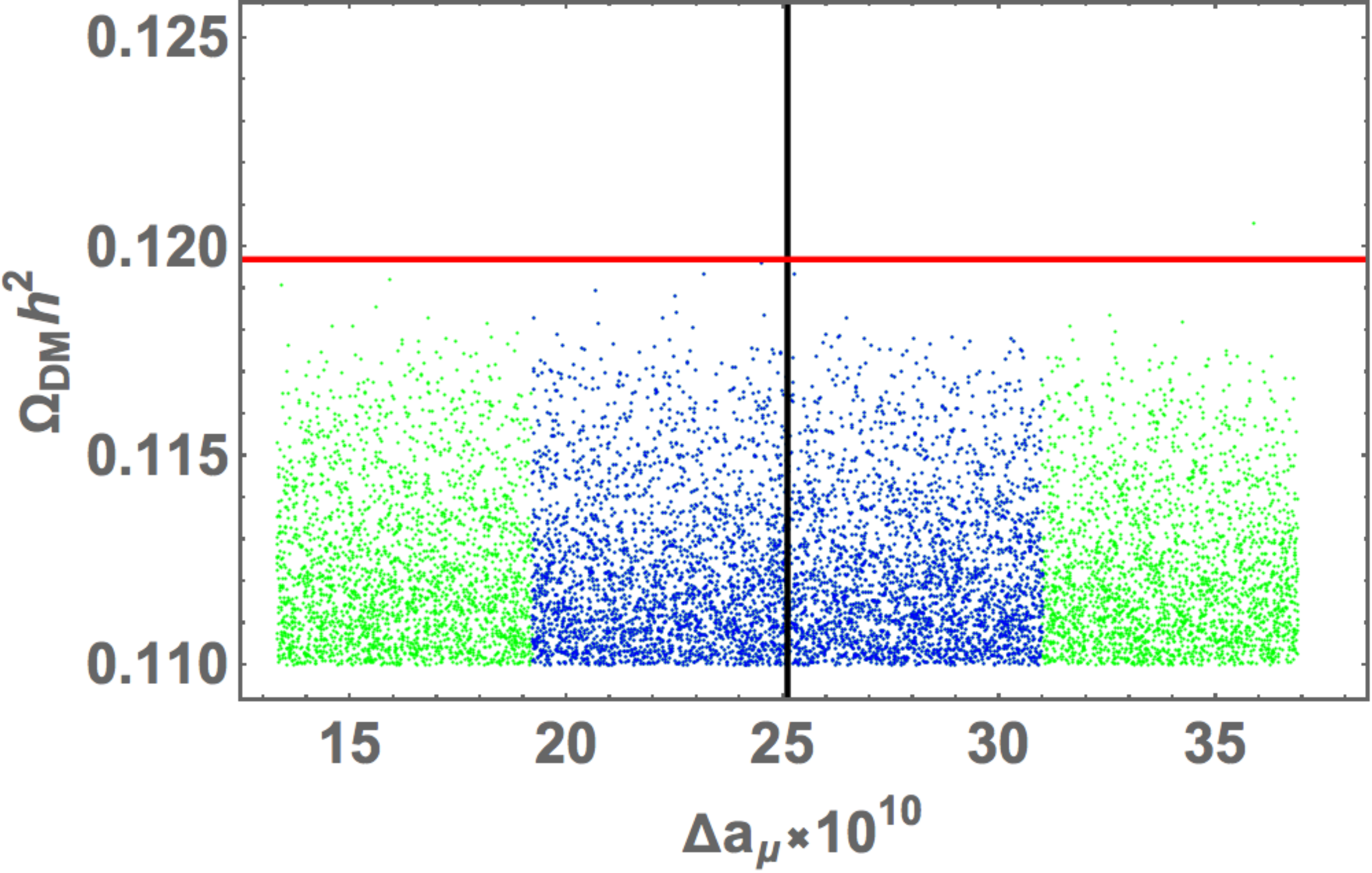}
 \caption{(Left) Allowed region in terms of $(\sigma v)_{\rm FL}\times10^{26}{\rm cm^3/sec}$ and $\Delta a_\mu\times 10^{10}$.
(Right) Preferred region for $(\Omega_{\rm DM}h^2, \Delta a_\mu)$. All of the points are satisfied with Fermi-LAT GeV excess within $1\sigma$. The red horizontal line indicates the central value of the DM relic abundance observed by Planck \cite{Planck:2015fie}.
In both left and right figures, the blue plot represents 1$\sigma$ interval of $\Delta a_\mu$ while the green one 2$\sigma$.
the black vertical line is the best fit value of $\Delta a_\mu$. }
 \label{fig:dam-fl}
\end{figure}
In the left panel of Fig.~\ref{fig:dam-fl}, we show the allowed region in terms of $(\sigma v)_{\rm FL}\times10^{26}{\rm cm^3/sec}$ and $\Delta a_\mu\times 10^{10}$.
The blue plot represents 1$\sigma$ interval of $\Delta a_\mu$ while the green one 2$\sigma$.
The black vertical line is the best fit value of $\Delta a_\mu$.
The figure suggests that we do not have allowed region within 1$\sigma$ of $(\sigma v)_{\rm FL}$ but have within 2$\sigma$.
In the right panel of Fig.~\ref{fig:dam-fl}, we show the allowed region in terms of $\Omega_{\rm DM} h^2$ and $\Delta a_\mu\times 10^{10}$.
This figure implies that most of the region is below the BF of $\Omega_{\rm DM} h^2$ but a few points reach at this value.

\section{Conclusions and discussions} \label{cons}
We have shown the successful explanation of Fermi-LAT GeV excess where DM is supposed to be muon specific, applying a modular $A_4$ symmetry. Thanks to nature of the symmetry, 
our DM interacts with pairs of muon only.
Moreover, we have explained sizable muon anomalous magnetic dipole moment without suffering from lepton flavor violations.
We have numerically demonstrated our allowed spaces on each the measurements above, randomly scanning our input parameters.

\vspace{0.5cm}
\hspace{0.2cm} 

\begin{acknowledgments}
The work of J.K. is supported in part by Korea Institute for Advanced Study (KIAS) Individual Grant. No. PG074202.
The work of H.O. was supported by the Junior Research Group (JRG) Program at the Asia-Pacific Center for Theoretical
Physics (APCTP) through the Science and Technology Promotion Fund and Lottery Fund of the Korean Government and was supported by the Korean Local Governments-Gyeongsangbuk-do Province and Pohang City.
\end{acknowledgments}

\bibliographystyle{utphys}
\bibliography{FL-mug2.bib}
\end{document}